\documentclass[reprint, aps,prl,twocolumn,superscriptaddress]{revtex4-1}
\usepackage{amssymb}
\usepackage{amsmath}
\usepackage{epsfig}
\usepackage{color}
\usepackage{graphics, graphicx}
\usepackage{bbold}
\usepackage{psfrag}
\usepackage{mathcomp}
\usepackage{subfigure}
\usepackage{verbatim}
\usepackage[colorlinks, citecolor=blue]{hyperref}
\usepackage[normalem]{ulem}
\def\cp#1{\mathbf{#1}}
\begin{document}

\title{Droplet under confinement: Competition and coexistence with soliton bound state}
\author{Xiaoling Cui}
\email{xlcui@iphy.ac.cn}
\affiliation{Beijing National Laboratory for Condensed Matter Physics, Institute of Physics, Chinese Academy of Sciences, Beijing 100190, China}
\affiliation{Songshan Lake Materials Laboratory, Dongguan, Guangdong 523808, China}
\author{Yinfeng Ma}
\affiliation{Beijing National Laboratory for Condensed Matter Physics, Institute of Physics, Chinese Academy of Sciences, Beijing 100190, China}
\affiliation{School of Physical Sciences, University of Chinese Academy of Sciences, Beijing 100049, China}
\date{\today}

\begin{abstract}
We study the stability of quantum droplet and its associated phase transitions in ultracold Bose-Bose mixtures uniformly confined in quasi-two-dimension with periodic boundary condition.
We show that the confinement-induced boundary effect can be significant when  increasing the atom number or reducing the confinement length, which destabilizes the quantum droplet towards the formation of a soliton bound state that has no density modulation along the confined direction. 
In particular, as increasing the atom number we find the soliton reentrance, while the droplet is stabilized only within a finite number window that sensitively depends on the confinement length. Near the droplet-soliton transitions, they can coexist with each other as two local minima in the energy landscape. Finally we map out the phase diagram for droplet-soliton transition and coexistence in the parameter plane of atom number and confinement length for $^{39}$K boson mixtures. %The intriguing competition between quantum droplet and soliton under uniform confinement may be probed in cold atoms experiments. 
\end{abstract}
\maketitle

{\it Introduction.}
Quantum droplet describes a self-bound many-body state that is stabilized by quantum effect. It has intrigued great attention recently in the field of ultracold atoms, given its successful observation in dipolar gases\cite{Pfau_1,Pfau_2,Pfau_3,Ferlaino,Modugno,Pfau_4,Ferlaino_2} and alkali Bose-Bose mixtures\cite{Tarruell_1,Tarruell_2,Inguscio,Modugno_2}. These dilute droplets, as pointed out in a pioneer work by Petrov\cite{Petrov}, are stabilized by a subtle balance between the mean-field attraction and the Lee-Huang-Yang(LHY) repulsion from quantum fluctuations. Similar stabilization mechanism has been extended to other  droplet systems including Bose-Fermi mixtures\cite{Cui, Adhikari, Rakshit1, Rakshit2, Wenzel,Yi} and dipolar mixtures\cite{Blakie,Santos_2}.

The stability of quantum droplet depends crucially on the geometry. In three-dimension(3D), the quantum pressure can dissociate the droplet at small atom number and lead to the liquid-gas transition as observed in experiments\cite{Pfau_1,Pfau_2,Pfau_3,Ferlaino,Modugno,Pfau_4,Ferlaino_2, Tarruell_1,Tarruell_2,Inguscio,Modugno_2}. In 2D and 1D, quantum droplet can be supported in quite different interaction regimes as compared to 3D, due to distinct LHY corrections\cite{Petrov_2}. In this context, it is conceptually important and also practically meaningful to investigate the confinement effect to droplet stability, which can bridge different droplet physics between different geometries. Previously, a few theoretical studies have revealed the significant change of LHY correction in quasi-low dimensions\cite{Santos, Jachymski, Zin, Buchler}.  In particular, it was shown that the LHY energy of alkali bosons can gradually change sign to negative as strengthening the confinement\cite{Zin, Buchler}, while the resulted instability of droplet and its associated transitions during the dimensional reduction have not been discussed therein. %Since the low-D atomic gas is generally reduced from 3D by confinements, it will be practically meaningful to explore the droplet properties during the realistic dimensional reduction, which will also help to understand different behaviors of droplets in different dimensions. The task is quite challenging, as it requires a thorough study of the interplay between confinement, quantum pressure and fluctuation effect.

\begin{figure}[t]
\includegraphics[width=9cm]{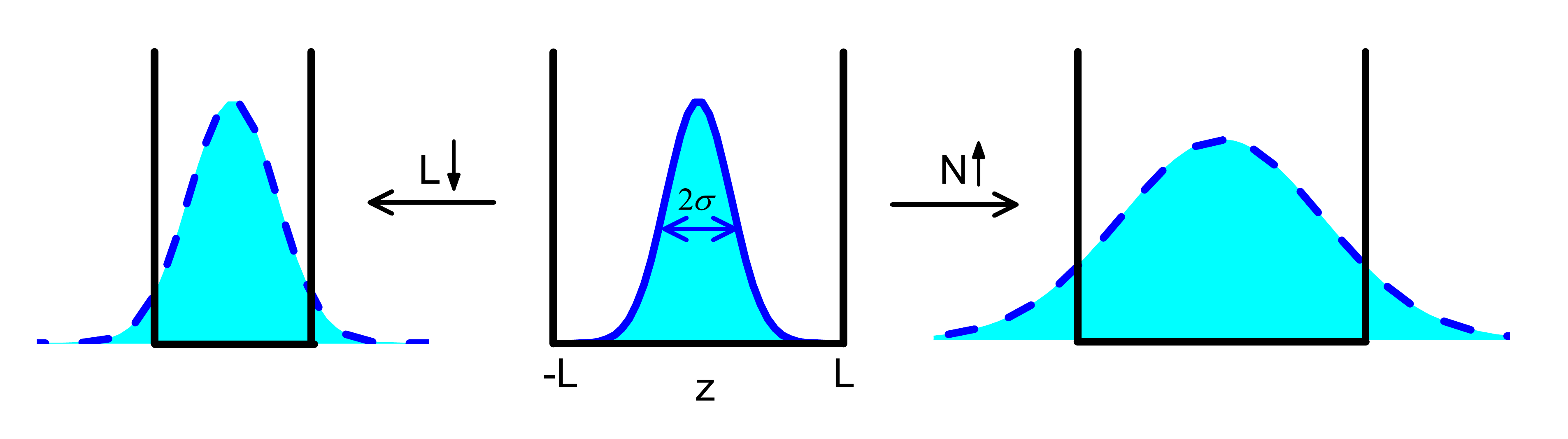}
\caption{Boundary effect to quantum droplet. Starting from a 3D droplet in a uniform trap with length much larger than the droplet size $L\gg \sigma$ (central plot), the boundary effect can become significant either by increasing the atom number (right) or by reducing the trap length (left). In both cases, we have $L\lesssim \sigma$ and the droplet will encounter instability.  Here the droplet wave function is plotted only along the trap direction.} \label{fig_schematic}
\end{figure}

Apart from the significant change of LHY correction, we note that the confinement can affect the droplet stability in two other non-trivial ways:

First, it introduces the boundary effect. As illustrated in Fig.\ref{fig_schematic}, for a droplet cloud confined uniformly with well-defined boundaries(central plot), the boundary effect can become significant when the droplet size $\sigma$ is comparable to the trap length $L$, either by increasing atom number $N$ or by reducing $L$. In either case, the droplet will adjust itself to be compatible with the boundary, which naturally causes instability. %Secondly, by changing the single-particle physics, the confinement can greatly modify the quantum pressure and LHY energy ($E_{\rm LHY}$), thereby significantly affecting the droplet formation. In particular, $E_{\rm LHY}$ can even change sign for strong confinements, as pointed out in Refs.\cite{Santos, Zin, Buchler}. Finally
Second, the confinement can introduce another channel of bound state to compete with the droplet. A well known example is the bright soliton in quasi-1D(q1D) that is stabilized by quantum pressure and mean-field attraction\cite{soliton_theory, soliton_expt1, soliton_expt2}. In a recent experiment, the droplet-soliton transition was explored in harmonically trapped quasi-1D Bose-Bose mixtures\cite{Tarruell_2}, while the confinement effect to qualitative change of LHY correction was not considered therein. %the experiment\cite{Tarruell_2} was conducted in 3D regime without considering the confinement effect to qualitative change of the LHY energy.

In this work, by fully taking into account the confinement effect,  we study the stability of quantum droplet and its associated transitions in Bose-Bose mixtures confined in q2D. To clearly see the boundary effect, we consider the uniform trap as depicted in Fig.\ref{fig_schematic} with periodic boundary condition. %, which has become
% with periodic boundary condition. In cold atoms, the uniform trap has become
%experimentally accessible with a tunable length\cite{box_trap1, box_trap2, box_trap3, box_trap4, box_trap5}. 
We find that when the boundary effect becomes significant, the droplet becomes unstable and gives way to a soliton bound state that displays no density modulation along the confined direction. %Different from the q1D soliton, here the formation of q2D soliton crucially relies on the contribution from quantum fluctuations.
As increasing $N$, a soliton to droplet transition occurs at relatively small $N$, while the boundary effect leads to the soliton reentrance at a larger $N$ and thus the droplet can only be stabilized within a finite number window that sensitively depends on the trap length. Near the droplet-soliton transitions, they can coexist with each other as two local minima in the energy landscape. Take the $^{39}$K boson mixture for example, we have analyzed in detail the droplet-soliton competition and finally mapped out the phase diagram for their transition and coexistence. 

{\it Model.} The Hamiltonian we consider for Bose-Bose mixture is $H=\int d{\bf r} H({\bf r})$, where ($\hbar=1$)
\begin{eqnarray}
H({\bf r})=\sum_{i=1,2}\Psi_i^{\dag}({\bf r})(-\frac{\nabla^2}{2m_i})\Psi_i({\bf r})+\sum_{ij}\frac{g_{ij}}{2} \Psi_i^{\dag}\Psi_j^{\dag}\Psi_j\Psi_i({\bf r}).  \nonumber
\end{eqnarray}
Here ${\bf r}=(x,y, z)$ is the coordinate; $m_i$ and $\Psi_i$ are respectively the mass and field operator of boson species $i$; $g_{ii}=4\pi a_{ii}/m_i$ and $g_{12}=2\pi a_{12}/\mu$ ($\mu=m_1m_2/(m_1+m_2)$) are the intra- and inter-species couplings. Given the atoms confined uniformly within $z\in [-L,L]$ and under periodic boundary condition, the momentum along $z$ are quantized as $k_n=n\pi/L$ $(n=0,\pm1,...)$. Based on the Bogoliubov theory for a homogeneous mixture with densities $n_1,n_2$\cite{book}, we can get the LHY energy per volume as:% ${\cal E}_{\rm LHY}\equiv E_{\rm LHY}/V$: % as the function of boson densities $n_1, \ n_2$:
\begin{eqnarray}
{\cal E}_{\rm LHY}&=&\int \frac{d^2 {\cp q}}{2(2\pi)^2} \frac{1}{2L} \sum_{n} \left[E^{+}_{n {\cp q}}+E^-_{n {\cp q}}-\sum_{i=1,2}(\epsilon^{(i)}_{n{\cp q}}+g_{ii}n_i) \right] \nonumber\\
&& + \int \frac{d^3 {\cp k}}{2(2\pi)^3} \frac{m_1g_{11}^2n_1^2+m_2g_{22}^2n_2^2+4\mu g_{12}^2n_1n_2}{{\cp k}^2}, \label{e_LHY}
\end{eqnarray}
Here ${\cp q}$ and ${\cp k}$ are respectively 2D and 3D momentum vectors, and the quasi-particle energies read
\begin{equation}
E^{\pm}_{n{\cp q}}=\sqrt{\frac{\omega_1^2+\omega_2^2}{2}\pm\sqrt{(\frac{\omega_1^2-\omega_2^2}{2})^2+4g_{12}^2n_1n_2 \epsilon^{(1)}_{n{\cp q}}\epsilon^{(2)}_{n{\cp q}}}} \label{mode}
\end{equation}
with $\omega_i=\sqrt{\epsilon^{(i)}_{n{\cp q}}(\epsilon^{(i)}_{n{\cp q}}+2g_{ii}n_i)}$ and $\epsilon^{(i)}_{n{\cp q}}=[(n\pi/L)^2+{\cp q}^2]/(2m_i)$. %In comparison to the previous studies on
We note that the LHY energy in quasi-low D was studied previously with different techniques aiming at the equal-mass mixtures\cite{Zin, Buchler}. In comparison, our scheme can apply for an arbitrary mass ratio.  For the equal-mass case, we have checked that Eq.\ref{e_LHY} can reproduce the LHY energy in effectively 2D\cite{Zin} or 3D\cite{Petrov} limit, given the boson densities are small or large.

To investigate the stability of self-bound state, we have to go beyond the bulk description and employ a spatially varying ansatz $\Psi_i({\bf r})$. Using the single-mode approximation  $\Psi_i({\bf r})=\sqrt{N_i}\phi({\bf r})$, we get the energy functional
\begin{equation}
E=E_{\rm kin}+E_{\rm mf}+E_{\rm LHY}, \label{E}
\end{equation}
with $E_{\rm kin}=\sum_i N_i \int d{\bf r} \phi^*({\bf r})(-\frac{\nabla^2}{2m_i})\phi({\bf r})$, $E_{\rm mf}=(g_{11}N_1^2/2+g_{22}N_2^2/2+g_{12}N_1N_2)\int d{\bf r} |\phi({\bf r})|^4$ and $E_{\rm LHY}=\int d{\bf r} {\cal E}_{\rm LHY}(n_i({\bf r}))$, where $n_i({\bf r})=N_i|\phi_i({\bf r})|^2$. We further assume the number ratio as $N_1/N_2=\sqrt{g_{22}/g_{11}}$ in order to minimize $E_{\rm mf}$\cite{Petrov}. Above assumptions have been shown to well predict the liquid-gas transition in  3D droplets\cite{Tarruell_1}. % (without trap) by comparing to the results from full simulation of coupled Gross-Pitaevskii equations.
 For the current  case with a uniform trap under periodic condition $\phi({\bf r})=\phi({\bf r}+2L{\bf e}_z)$, we adopt an extended Gaussian-type ansatz as follows:
\begin{equation}
\phi({\bf r})=\frac{1}{\sqrt{{\cal N}}} {\rm exp}(-\frac{x^2+y^2}{2\sigma_{xy}^2}) \left[\sum_{\nu=-\infty}^{\infty} {\rm exp}\left(-\frac{(z-2\nu L)^2}{2\sigma_{z}^2}\right) \right]. \label{wf}
\end{equation}
Here ${\cal N}$ is the normalization factor; $\sigma_{xy}$ and $\sigma_z$ are two variational parameters and represent, respectively, the sizes of bound state along $xy$ and $z$. The ground state can be obtained by minimizing the energy functional (\ref{E}) in terms of $\sigma_{xy}$ and $\sigma_z$.

%By searching for the energy minimum in terms of $\sigma_{xy}$ and $\sigma_z$ (i.e., $\partial E/\partial\sigma_{xy,z}=0$ and $\partial^2E/\partial\sigma^2_{xy,z}>0)$), we find two possible candidates for the ground state:  a droplet state if both $\sigma_{xy}$ and $\sigma_z$ are finite, and a soliton-like state if $\sigma_{xy}$ is finite and $\sigma_z\rightarrow\infty$ (no density modulation along $z$). Different from free space case ($L=\infty$), here no gaseous ground state ($\sigma_{xy,z}\rightarrow\infty$)  can be found with a finite $L$.

In this work, we specifically consider the two hyperfine states of $^{39}$K atoms, $|1\rangle\equiv|F=1,m_F=0\rangle,\ |2\rangle\equiv|F=1,m_F=-1\rangle$, as have been well studied in 3D droplet experiments\cite{Tarruell_1,Tarruell_2,Inguscio}. In this case,  $a_{22}=35a_B,\ a_{12}=-53a_B$ ($a_B$ is the Bohr radius), and $a_{11}$ is highly tunable by magnetic field. %across the mean-field collapse value $a_{11,c}=80.3a_B$.
We will focus on the mean-field collapse regime with $\delta a\equiv a_{12}+\sqrt{a_{11}a_{22}}<0$ and study how the uniform confinement affects the quantum droplet. As we consider small $|\delta a|\ (\ll a_{11},a_{22},|a_{12}|)$, in calculating $E_{\rm LHY}$ we make the approximation $\delta a=0$  to avoid the phonon instability due to complex spectrum (\ref{mode}). Other rectified theories on this have appeared recently\cite{HuHui, Ota, YinLan}.  Throughout the paper, we choose the length unit as $l_0=1\mu m$ and the energy unit as $E_0=1/(2ml_0^2)$, with mass $m\equiv m_1=m_2$ for $^{39}$K atoms.

{\it Results.} By searching for the energy minimum in terms of $\sigma_{xy}$ and $\sigma_z$, i.e., $\partial E/\partial\sigma_{xy,z}=0$ and $\partial^2E/\partial\sigma^2_{xy,z}>0$, we find two candidates for the ground state:  one is with finite $\sigma_{xy}$ and finite $\sigma_z$, which is smoothly connected to the 3D droplet for large $L$ and is thus referred as {\it droplet}; the other is with finite $\sigma_{xy}$ and $\sigma_z\rightarrow\infty$, which exists only under confinement and is referred as {\it soliton}. Different from free space case, here no gaseous ground state (both $\sigma_{xy,z}\rightarrow\infty$)  can be found for finite $L$.

\begin{figure}[h]
\includegraphics[width=8cm, height=11cm]{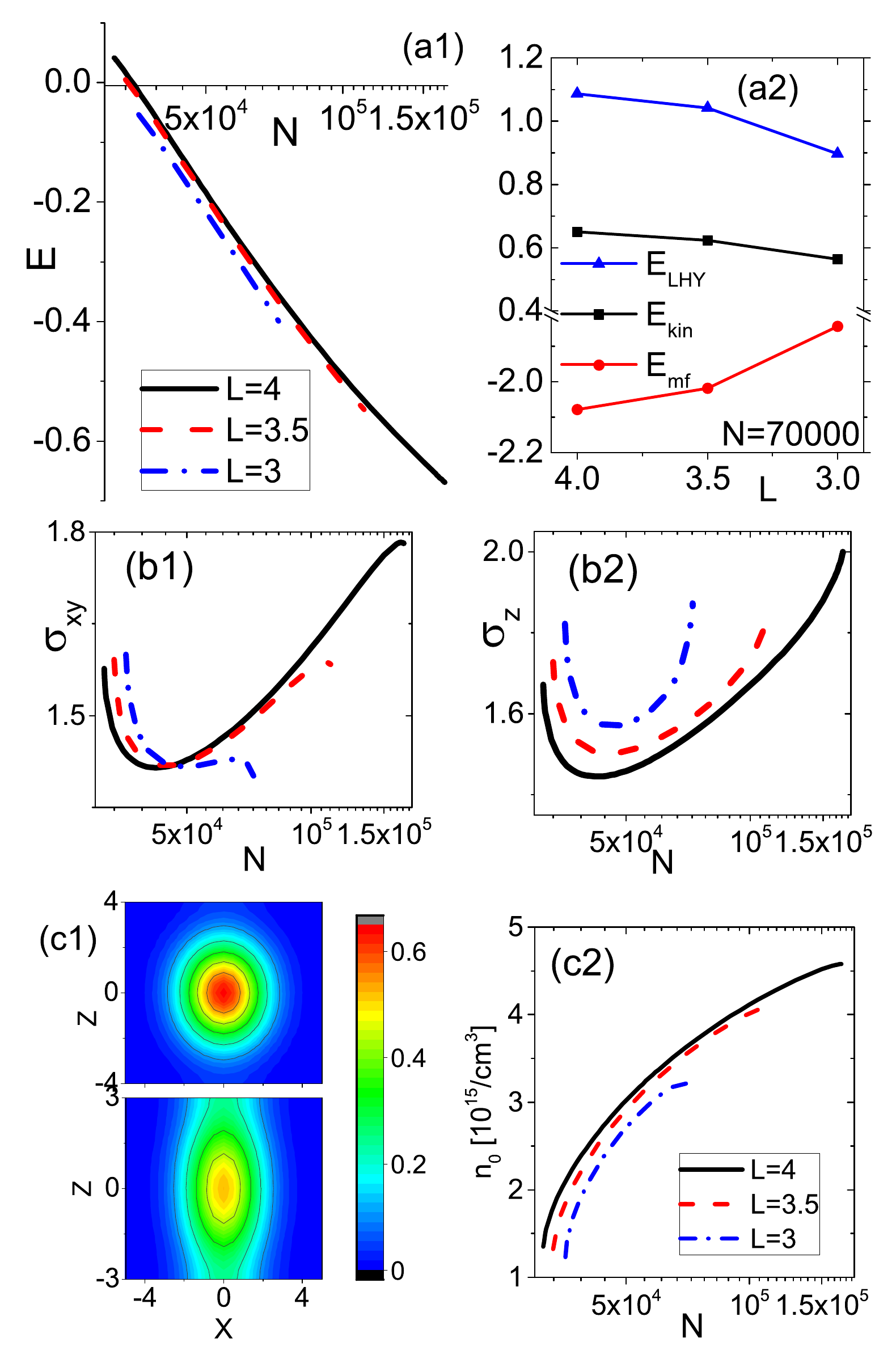}
\caption{Droplet state of $^{39}$K atoms in q2D with $\delta a=-5a_B$.
(a1)Droplet energy $E$ as a function of atom number $N$ at different $L=4,\ 3.5,\ 3$. (a2) $E_{\rm LHY}$, $E_{\rm kin}$ and $E_{\rm mf}$ at different $L$ for a given $N=7\times10^4$.  (b1,b2) Droplet sizes $\sigma_{xy}$ and $\sigma_z$ as functions of $N$ at various $L$ (with the same line style as in (a1)). (c1) Contour plot of droplet wave function $\phi({\bf r})$ in (x,z) plane (with $y=0$) for a given $N=7\times10^4$ at $L=4$ (upper panel) and $3$ (lower panel). (c2) Peak density $n_0$ (in unit of $10^{15}/{\rm cm}^3$) as a function of $N$. Here the length and energy units are respectively $l_0=1\mu m$ and $E_0=1/(2ml_0^2)$.} \label{fig_droplet}
\end{figure}

{\it (I) Droplet solution.}
Fig.\ref{fig_droplet} shows the droplet solution as varying $N$ at different   $L$. One can see from Fig.\ref{fig_droplet}(a1) that the total energy of droplet continuously decreases as shrinking $L$, which can be attributed to the reduced kinetic and LHY energies, as shown by $E_{\rm kin}$ and $E_{\rm LHY}$ in Fig.\ref{fig_droplet}(a2). The reduction of $E_{\rm kin}$ is due to the enlarged energy gap and thus the suppressed excitation along $z$, while the reduction of $E_{\rm LHY}$ here is consistent with that found in Ref.\cite{Zin, Buchler}.  Another remarkable effect of finite $L$ is that, now the droplet only survives within a finite number window $[N_{d1},N_{d2}]$, unlike the free space droplet that just requires a lower number bound. This number window becomes narrower for smaller $L$, due to the existence of another competitive bound state (soliton, as discussed later). In particular, we see that a small $L$ also gives rise to a small upper bound $N_{d2}$, which is consistent with the boundary effect as illustrated in Fig.\ref{fig_schematic}.

Fig.\ref{fig_droplet}(b1,b2) show both $\sigma_{xy}$ and $\sigma_z$ evolving non-monotonically with $N$. Near the vanishing point of droplet ($N\sim N_{d2}$), shrinking $L$ will lead to a smaller $\sigma_{xy}$ but a larger $\sigma_z$. This means that by tightening the confinement, %more weight of the droplet transfers from the free ($xy$) to confined ($z$) direction; accordingly, 
the droplet wave function will change from isotropic  to highly elongated shape (along $z$), as shown in Fig.\ref{fig_droplet}(c1). This counter-intuitive change share the same reason with the suppressed $E_{\rm kin}$ as shown in Fig.\ref{fig_droplet}(a2), i.e., the lower $E_{\rm kin}$ corresponds to an extended density distribution  along $z$. %, can be attributed to the enlarged energy gap along $z$, which tends to froze the system to the ground state mode along $z$ and thus the wave function becomes more elongated in this direction.
Because of such extended distribution, the peak density of the droplet $n_0$ also decreases as $L$ shrinks, see Fig.\ref{fig_droplet}(c2). Here $n_0\sim 10^{15}/{\rm cm}^3$, the same order as the typical density in 3D droplet experiment\cite{Inguscio}.

\begin{figure}[h]
\includegraphics[width=6.5cm]{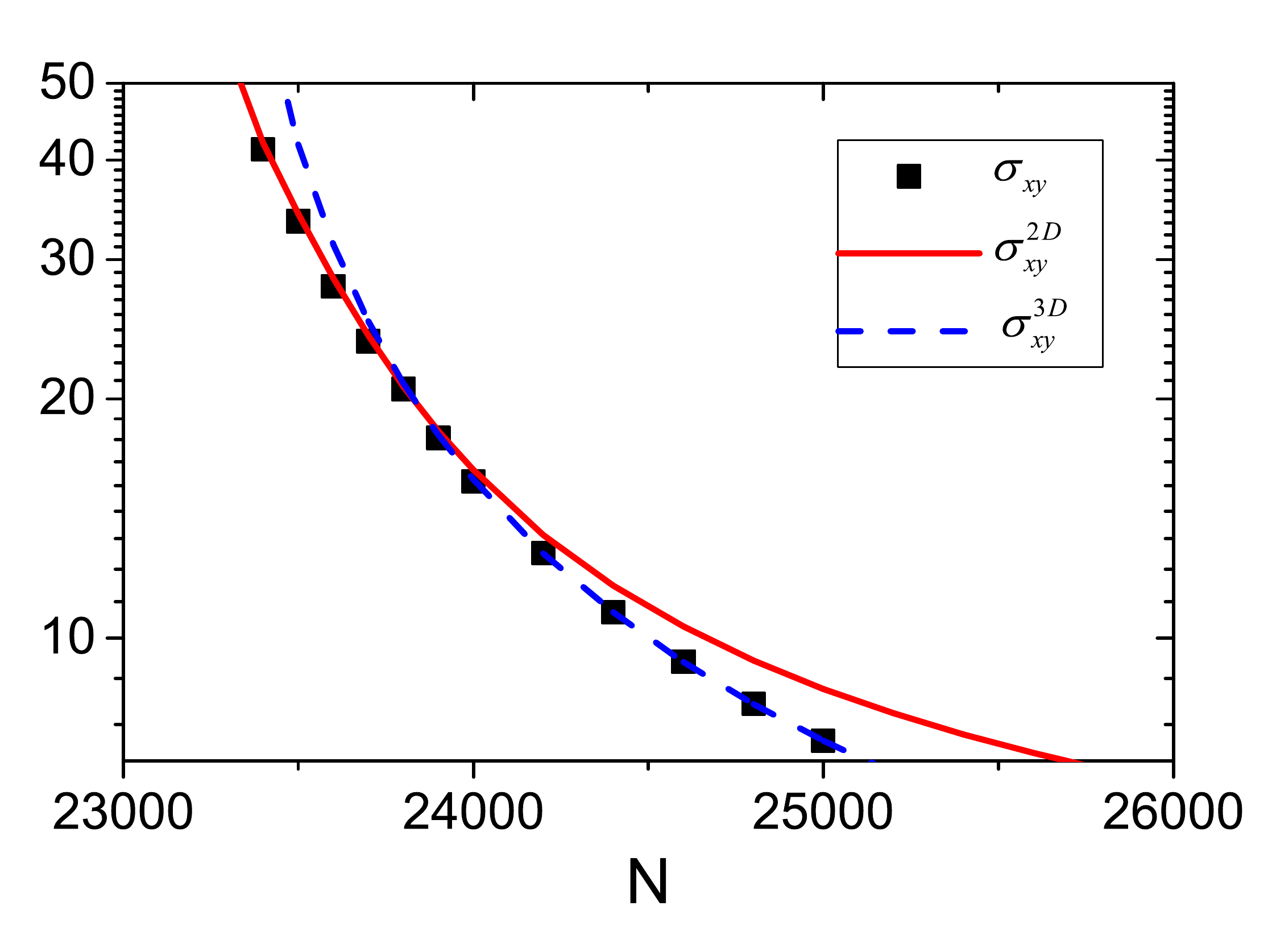}
\caption{Transverse size of soliton state, $\sigma_{xy}$(black square), for $^{39}$K atoms in q2D.  Here $\delta a=-5a_B$ and $L=3$. Red solid and blue dashed lines are the fit to results in $2D$ and $3D$ limits (see text). The length unit is $l_0=1\mu m$.} \label{fig_soliton}
\end{figure}

{\it (II) Soliton solution.} The uniform trap along $z$ can also support a soliton bound state ($E<0$) at $\sigma_{z}=\infty$ and a finite $\sigma_{xy}$, where the density modulation is allowed only along free ($xy$) directions but not along the confined ($z$) direction.  Different from the q1D soliton\cite{soliton_theory, soliton_expt1, soliton_expt2}, here only the kinetic and mean-field terms are inadequate to support the q2D soliton. For instance, in the 2D limit we have $E_{\rm kin}\sim \sigma_{\rm xy}^{-2}$ and $E_{\rm mf}\sim -L^{-1}\sigma_{\rm xy}^{-2}$, and one has to incorporate the contribution from $E_{\rm LHY}$ to enable an energy minimum at finite $\sigma_{xy}$. %, so the energy minimum is either at $\sigma_{\rm xy}=0$(collapse) or at $\sigma_{\rm xy}=\infty$(gas). In this way, one has to taken into account the contribution from $E_{LHY}$ to stabilize the soliton bound state.
It is noted that such LHY-stabilized soliton in 2D limit is equivalent to the 2D droplet studied in Ref.\cite{Petrov_2}.

In Fig.\ref{fig_soliton} we show the soliton size $\sigma_{xy}$ as varying $N$ at fixed $L=3\mu m$. To analyze its behavior in the limits of small and large $N$, we utilize the analytical expressions of $\epsilon_{\rm LHY}$ in 2D and 3D limits (same as Eq.(7) in \cite{Zin} and Eq.(5) in \cite{Petrov}), and obtain the integrated LHY energy as 
\begin{eqnarray}
E_{\rm LHY}^{\rm 2D}&=&\frac{2D^2}{m\sigma_{xy}^2} \left(\ln [\frac{4{L}D^{1/2}}{{\sigma}_{xy}} ] + \frac{8{L}^2}{9{\sigma}_{xy}^2} D \right); \\
E_{\rm LHY}^{\rm 3D}&=&\frac{1024 {L} }{75\pi m{\sigma}_{xy}^3}  D^{5/2}
\end{eqnarray}
with $D=(N_1a_{11}+N_2a_{22})/(2L)$. Minimizing the total energy $E$, we can obtain the equilibrium size, $\sigma^{\rm 2D}_{xy}$ or $\sigma^{\rm 3D}_{xy}$, in 2D or 3D limit.  Fig.\ref{fig_soliton} shows that $\sigma^{\rm 2D}_{xy}$ ($\sigma^{\rm 3D}_{xy}$) fits well to the soliton size $\sigma_{xy}$ in small (large) $N$ limit. 

\begin{figure}[h]
\includegraphics[width=9cm]{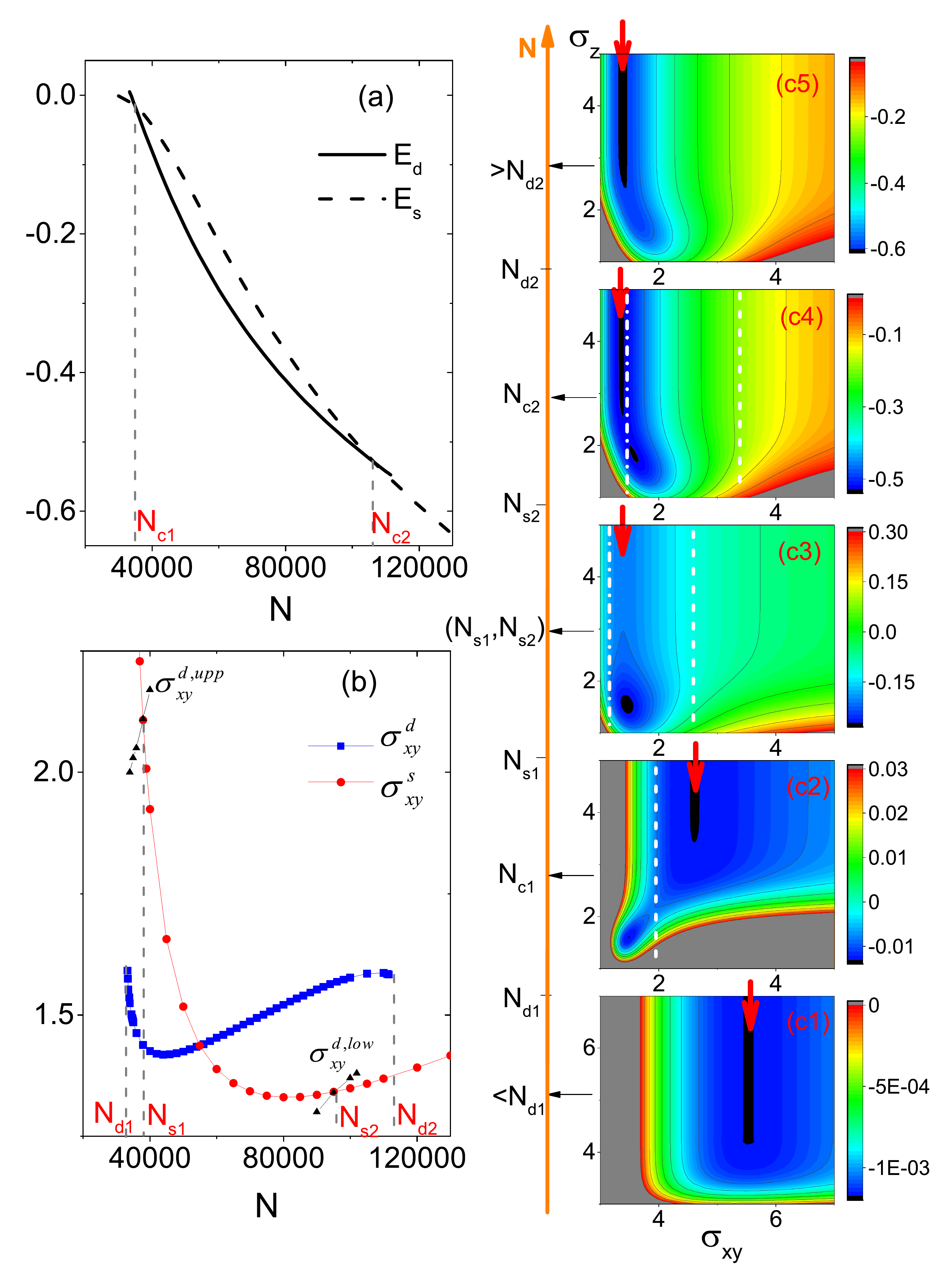}
\caption{Droplet-Soliton transition and coexistence at $\delta a=-5 a_B$ and $L=3.5 \mu m$. (a) Energies of droplet ($E_d$) and soliton ($E_s$) as functions of $N$. The energy crossings determine two transition points at $N_{c1}$ and $N_{c2}$. (b) Transverse sizes of droplet ($\sigma_{xy}^{d}$) and soliton ($\sigma_{xy}^{s}$), in comparison with $\sigma_{xy}^{d,low}$ and $\sigma_{xy}^{d,upp}$ defining the droplet region (see text). The droplet is locally stable for $N\in(N_{d1},N_{d2})$, and is the only stable (ground) state for $N\in(N_{s1},N_{s2})$ when the soliton enters droplet region. Droplet-soliton coexistence occurs at $N\in(N_{d1},N_{s1})\bigcup(N_{s2},N_{d2})$.  (c1-c5) Contour plot of $E(\sigma_{xy}, \sigma_{z})$ for various $N(10^4)$:  $3 (c1),\ 3.48(=N_{c1},c2),\ 6 (c3),\ 10.66(=N_{c2},c4),\ 11.5(c5)$. The white dashed-dot and dashed lines mark the locations of $\sigma_{xy}^{d,low}$ and $\sigma_{xy}^{d,upp}$, and the red arrows mark $\sigma_{xy}^s$. The length and energy units are the same as Fig.\ref{fig_droplet}.} \label{fig_transition}
\end{figure}

{\it (III) Droplet-soliton transition and coexistence.} After identifying the individual property of droplet and soliton, now we turn to their competition. In Fig.\ref{fig_transition}, we show their transition and coexistence as varying $N$ for a fixed $L=3.5\mu m$. As seen from Fig.\ref{fig_transition}(a), the energies of droplet and soliton cross twice as increasing $N$, which gives two transition points respectively at $N_{c1}$ and $N_{c2}$. Their individual stability and mutual competition can be clearly seen from the energy contour plots $E(\sigma_{xy},\sigma_z)$ in Fig.\ref{fig_transition}(c1-c5), together with the comparison of their transverse sizes $\sigma_{xy}$ shown  in Fig.\ref{fig_transition}(b).

For small $N$, the only energy minimum represents a soliton state, i.e., at $\sigma_z\rightarrow\infty$ and a finite $\sigma_{xy}$ (see Fig.\ref{fig_transition}(c1)). As increasing $N$ to $N_{d1}$, the droplet start to emerge as an additional energy minimum at finite $\sigma_z$ and a smaller $\sigma_{xy}$(Fig.\ref{fig_transition}(b)). The double minima reach the same energy at the first transition point  $N_{c1}$ (Fig.\ref{fig_transition}(c2)).

To facilitate later discussions, let us define the {\it droplet region} in the energy landscape along $\sigma_{xy}$, with lower bound $\sigma_{xy}^{d,low}$ and upper bound $\sigma_{xy}^{d,upp}$ (marked by the dashed-dot and dashed lines in Fig.\ref{fig_transition}(c2-c4)). Within this region, for any $\sigma_{xy}\in(\sigma_{xy}^{d,low},\sigma_{xy}^{d,upp})$, the energy minimum occurs at a finite $\sigma_z$. %Once the soliton enters this region, i.e., when its size 
By this definition, the droplet solution stays right within the droplet region, see Fig.\ref{fig_transition}(c2-c4). Then, if the soliton solution also lies in this region(see red arrow in Fig.\ref{fig_transition}(c3)), i.e., 
when $\sigma_{xy}^s\in (\sigma_{xy}^{d,low},\sigma_{xy}^{d,upp})$, the soliton will become locally unstable and flow from $\sigma_z=\infty$ to the droplet minimum. In Fig.\ref{fig_transition}(b), we denote the atom number at the intersection of $\sigma_{xy}^s$ and $\sigma_{xy}^{d,upp}$ ($\sigma_{xy}^{d,low}$)  as $N_{s1}$ ($N_{s2}$). Correspondingly, when $N\in[N_{s1},N_{s2}]$ the droplet is the only stable (ground) state, see Fig.\ref{fig_transition}(c3). For $N$ beyond $N_{s2}$, the soliton moves outside the droplet region and they can coexist again. Their second transition occurs at $N_{c2}$ when the two minima have the same energy, see Fig.\ref{fig_transition}(c4). The coexistence stops at $N=N_{d2}$ when the droplet solution disappears, and for $N>N_{d2}$ the only stable state becomes soliton again, see Fig.\ref{fig_transition}(c5).

%Above analysis shows that under confinement, the atom number can conveniently tune the competition and coexistence between droplet and soliton, through the manipulation of their individual energies and sizes. Importantly,
From above, we can see that %the stability of soliton can be strongly destroyed when it  affected by the dr
the droplet-soliton competition is most pronounced when the soliton enters the droplet region, or equivalently, when they have similar sizes along free ($xy$) directions. %For instance, when the soliton size enters the droplet region $\sigma_{xy}^s\in(\sigma_{xy}^{d,low}, \sigma_{xy}^{d,upp})$, the soliton will become locally unstable and easily flow to the droplet minimum with lower energy, see Fig.\ref{fig_transition}(c3).
On the other hand,  the instability of droplet as well as the reentrance of soliton at large $N>N_{d2}$ can be attributed to the boundary effect(Fig.\ref{fig_schematic}), when the droplet size along $z$ is comparable with $L$. For instance, at $N_{d2}$ we have $\sigma_z=1.86\mu m$, beyond half of $L(=3.5 \mu m)$.

\begin{figure}[h]
\includegraphics[width=7.5cm]{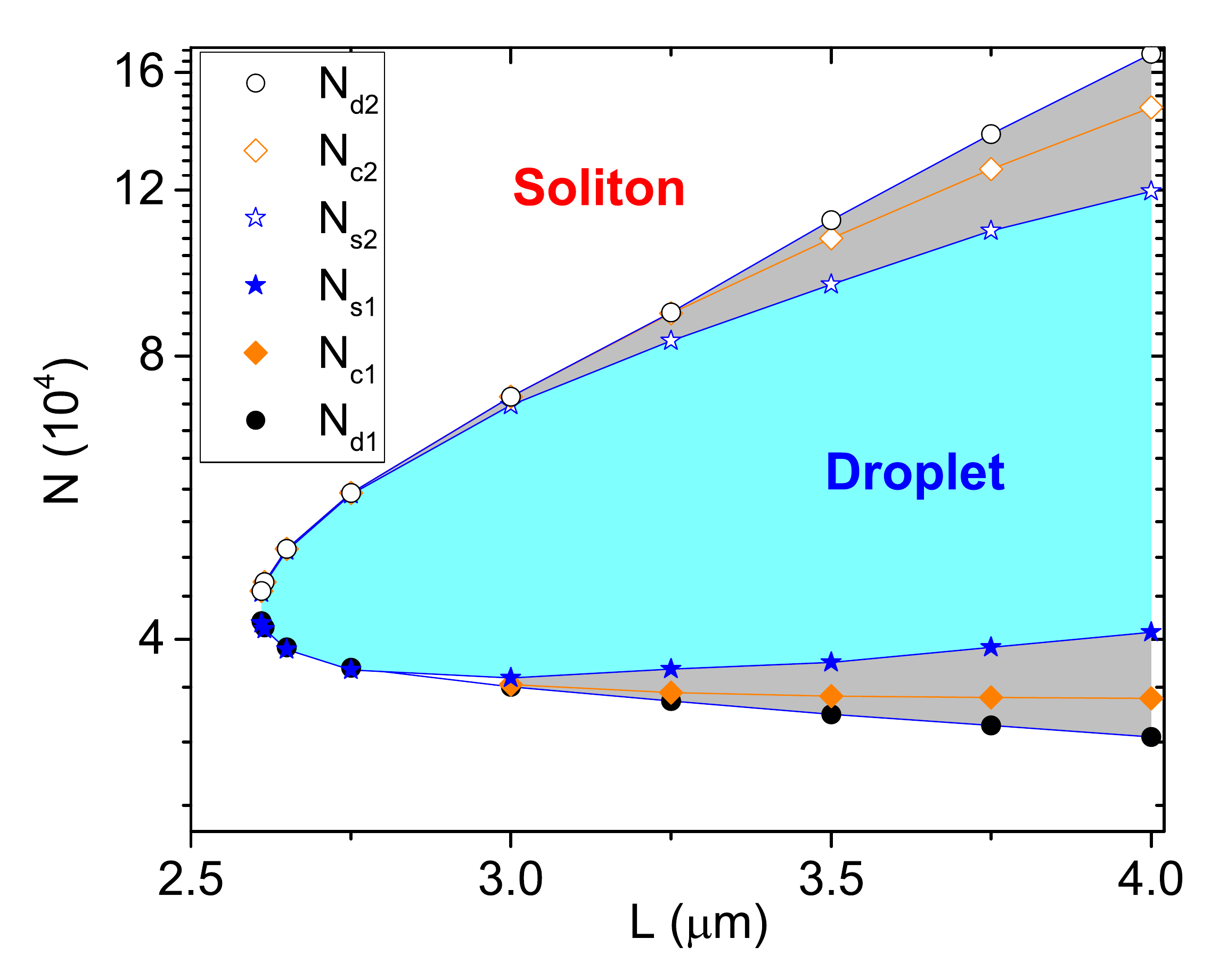}
\caption{Phase diagram in the $(N,L)$ plane for $^{39}$K mixture at $\delta a=-5a_B$.
%(b) is in term of atom number $N$ and $\delta a$ at a fixed trap length $L=3.5 \mu m$.
The droplet, soliton, and their coexistence regions are respectively shown by blue, white, and gray colors. Their phase boundaries are given by $N_{d1},\ N_{s1},\ N_{s2}$ and $N_{d2}$ (see text). Droplet-soliton transitions (energy crossing) occur at $N_{c1}$ and $N_{c2}$,  denoted by solid and hollow orange diamonds. %The global ground state is droplet for $N\in(N_{c1},N_{c2})$, and soliton otherwise.
} \label{fig_diagram}
\end{figure}

{\it (IV) Phase diagram. } To fully explore the confinement effect, we have carried out similar analysis for different $L$ and arrived at the phase diagram in the $(N,L)$ plane as shown in Fig.\ref{fig_diagram}. One can see that the droplet state (blue color) only survives within a finite number window that sensitively depends on the value of $L$. It will give way to the soliton state (white color) for very large or small $N$, or for small $L$. Near their transition points $N_{c1}$ and $N_{c2}$ (orange diamonds), the droplet and soliton can coexist with each other, and their coexistence region (gray color) also depends sensitively on $L$.

In fact, for $L\in(2.6,3)\mu m$ we find continuous transitions between droplet and soliton, i.e., the location of energy minimum continuously change between finite and infinite $\sigma_z$ across the phase boundaries, For $L<2.6\mu m$, no droplet solution can be found and the soliton is the only stable (ground) state. Again this can be attributed to the large energy gap along $z$, which rules out the possibility of density modulation in this direction. % physical picture for this finding is that, for very small $L$, the large energy gap along $z$ rules out the possibility of density modulation in this direction, and thus the soliton state is always more favored for any particle number.

%Fig.\ref{fig_diagram} can be readily detected in experiments.  The droplet and soliton states can be identified by measuring their density modulations along the confined direction, and their transitions can be explored through the discontinuous number change or droplet fragmentation when sweeping across the phase boundary, as recently conducted in experiment with a different setup\cite{Tarruell_2}.

{\it Discussion.}
%In summary, we have shown that the uniform confinement can induce rich physics of droplet-soliton competition and coexistence in Bose-Bose mixtures. Experimentally, the droplet and soliton states can be identified by measuring their density modulation along the confined direction, and their transitions can be explored through the discontinuous number change or droplet fragmentation when sweeping across the phase boundary, as recently conducted in experiment with a different setup\cite{Tarruell_2}.
In this work we have adopted the local density approximation(LDA) to compute $E_{\rm LHY}$, which was shown to predict the 3D droplets quantitatively well\cite{Tarruell_1, Inguscio}. Here we remark that %a natural question is that whether
the LDA is even more qualified in our case, especially along the confined direction with small $L$.  This is because as reducing $L$, the density distribution gets more extended along $z$ and the kinetic energy is further suppressed(Fig.\ref{fig_droplet}(a2,c1)). In fact, we have $\eta_z\equiv E_{\rm kin,z}/E_{\rm LHY}\ll 1$ in a broad parameter regime considered in this work. Take the case in Fig.\ref{fig_transition} for instance, the ratio $\eta_z$ is $0.46$ at $N_{c1}$ and gets even smaller to $0.08$ at $N_{c2}$. This is to say, the typical  length at which the density varies is visibly longer than that characterizing the LHY correction, which justifies the use of LDA in our setup.

%It is also noted that the formula and analysis in this work can be directly generalized to boson mixtures in q1D geometry and with unequal masses, where similar competition physics between droplet and soliton is expected to persist.

Though we have taken the periodic boundary condition, our results shed important light on the hard-wall boundary case as realized in current experiments\cite{box_trap1, box_trap2, box_trap3, box_trap4, box_trap5}. We expect the hard-wall boundary can equally cause the instability of quantum droplet at small $L$ or large $N$ (see Fig.\ref{fig_schematic}).  Nevertheless, in this case the droplet  cannot extend outside the boundary, in contrast with the periodic case (see Fig.\ref{fig_droplet}(c1)), and therefore the actual phase diagram need to be re-examined. 
Finally, it is worth to point out that the boundary effect here  is unlikely to apply for harmonic confinements, where the boundary cannot be clearly defined and the eigen-mode is also different. This follows that the physics near $N_{c2}$, as mostly driven by the boundary effect, would disappear for harmonically confined systems.   %Since the boundary effect essentially drives the droplet instability at large $N$, we expect that in a harmonic trap the physics near $N_{c2}$ will be largely modified while the physics near $N_{c1}$ is still robust. % less affected by the type of confinement.
This expectation is consistent with the recent experiment of harmonically trapped Bose-Bose mixtures in q1D\cite{Tarruell_2}, where only one droplet-soliton transition (corresponding to $N_{c1}$ in this work) was observed. %Our formula and analysis can be directly generalized to mixtures uniformly confined in q1D or with unequal masses, where similar competition physics between droplet and soliton is expected to persist.

\acknowledgments
{\bf Acknowledgment.}
The work is supported by the National Key Research and Development Program of China (2018YFA0307600, 2016YFA0300603), the National Natural Science Foundation of China (No.11534014, No.12074419), and the Strategic Priority Research Program of Chinese Academy of Sciences (No. XDB33000000).


\begin{thebibliography}{99}

%dipole system:
\bibitem{Pfau_1}I. Ferrier-Barbut, H. Kadau, M. Schmitt, M. Wenzel, and T. Pfau, Phys. Rev. Lett. {\bf 116}, 215301 (2016).
\bibitem{Pfau_2}M. Schmitt, M. Wenzel, F. B{\"{o}}ttcher, I. Ferrier-Barbut, and T. Pfau,  Nature {\bf 539}, 259 (2016).
\bibitem{Pfau_3}I. Ferrier-Barbut, M. Schmitt, M. Wenzel, H. Kadau, and T. Pfau, J. Phys. B {\bf 49}, 214004 (2016).
\bibitem{Ferlaino}L. Chomaz, S. Baier, D. Petter, M.J. Mark, F. W{\" a}chtler, L. Santos, and F. Ferlaino, Phys. Rev. X {\bf 6}, 041039 (2016).
%supersolid in dipolar droplet:
\bibitem{Modugno}L. Tanzi, E. Lucioni, F. Fam\`a, J. Catani, A. Fioretti, C. Gabbanini, R. N. Bisset, L. Santos, and G. Modugno, Phys. Rev. Lett. {\bf 122}, 130405 (2019).
\bibitem{Pfau_4}F. B{\" o}ttcher, J.-N. Schmidt, M. Wenzel, J. Hertkorn, M. Guo, T. Langen, and T. Pfau, Phys. Rev. X {\bf 9}, 011051 (2019).
\bibitem{Ferlaino_2}L. Chomaz, D. Petter, P. Ilzh{\"{o}}fer, G. Natale, A. Trautmann, C. Politi, G. Durastante, R.M.W. van Bijnen, A. Patscheider, M. Sohmen, M.J. Mark, and F. Ferlaino, Phys. Rev. X {\bf 9}, 021012 (2019).

%two-species bosons:
\bibitem{Tarruell_1}C.R. Cabrera, L. Tanzi, J. Sanz, B. Naylor, P. Thomas, P. Cheiney, and L. Tarruell, Science {\bf 359}, 301 (2018).
\bibitem{Tarruell_2}P. Cheiney, C. R. Cabrera, J. Sanz, B. Naylor, L. Tanzi, L. Tarruell,  Phys. Rev. Lett. {\bf 120}, 135301 (2018). %soliton to droplet transition
\bibitem{Inguscio}G. Semeghini, G. Ferioli, L. Masi, C. Mazzinghi, L. Wolswijk, F. Minardi, M. Modugno, G. Modugno, M. Inguscio, M. Fattori, Phys. Rev. Lett. {\bf 120}, 235301 (2018).
%\bibitem{Inguscio_2}G. Ferioli, G. Semeghini, L. Masi, G. Giusti, G. Modugno, M. Inguscio, A. Gallem\acute{i}, A. Recati, and M. Fattori, Phys. Rev. Lett. {\bf 122}, 090401 (2019).  %collision of droplet
\bibitem{Modugno_2}C. D'Errico, A. Burchianti, M. Prevedelli, L. Salasnich, F. Ancilotto, M. Modugno, F. Minardi, and C. Fort, Phys. Rev. Research {\bf 1}, 033155 (2019). %K-Rb droplet

%Alessia Burchianti, Chiara D'Errico, Marco Prevedelli, Luca Salasnich, Francesco Ancilotto, Michele Modugno, Francesco Minardi, Chiara Fort, Condens. Matter  5, 21 (2020). %K-Rb droplet



%first theory of droplet:
\bibitem{Petrov}D.S. Petrov, Phys. Rev. Lett. {\bf 115}, 155302 (2015).

%B-F droplet:
\bibitem{Cui}X. Cui, Phys. Rev. A {\bf 98}, 023630 (2018).
\bibitem{Adhikari}S. Adhikari, Laser Phys. Lett {\bf 15}, 095501 (2018).
\bibitem{Rakshit1}D. Rakshit, T. Karpiuk, M. Brewczyk, and M. Gajda, SciPost Phys {\bf 6}, 079 (2019).
\bibitem{Rakshit2}D. Rakshit, T. Karpiuk, P. Zin, M. Brewczyk, M. Lewenstein, and M. Gajda, New J. Phys. {\bf 21}, 073027 (2019). %B-F droplet in low-D
\bibitem{Wenzel}M. Wenzel, T. Pfau and I. Ferrier-Barbut, Physica Scripta {\bf 93}, 10 (2018). %fermion impurity in dipolar bosonic droplet
\bibitem{Yi}J.-B. Wang, J.-S. Pan, X. Cui, W. Yi, Chin. Phys. Lett. {\bf 37}, 076701 (2020).
%droplet in soc system:
%J. S\acute{a}nchez-Baena, J. Boronat, F.Mazzanti, arxiv:2007.04196.

%droplet in binary dipoles:
\bibitem{Blakie}Joseph C. Smith, D. Baillie, and P. B. Blakie, arxiv:2007.00366.
\bibitem{Santos_2}R. N. Bisset, L. A. Pe$\tilde{n}$a Ardila, and L. Santos, arxiv:2007.00404.


%in low-D:
\bibitem{Petrov_2}D. S. Petrov and G. E. Astrakharchik,  Phys. Rev. Lett. {\bf 117}, 100401 (2016).
%\bibitem{Nishida}Y. Sekino, Y. Nishida, Phys. Rev. A {\bf 97}, 011602(R) (2018). %1d with 3-body interaction
%quasi-low-D: dimensional crossover
%dipole gas:
\bibitem{Santos}D. Edler, C. Mishra, F. W{\" a}chtler, R. Nath, S. Sinha, and L. Santos, Phys. Rev. Lett. {\bf 119}, 050403 (2017). %dipole gas in quasi-1d
\bibitem{Jachymski}K. Jachymski and R. Oldziejewski, Phys. Rev. A {\bf 98}, 043601 (2018).
%1911.02384: unpublished
%B-B mixture:
\bibitem{Zin}P. Zin, M. Pylak, T. Wasak, M. Gajda, and Z. Idziaszek, Phys. Rev. A {\bf 98}, 051603(R) (2018).
\bibitem{Buchler}T. Ilg, J. Kumlin, L. Santos, D. S. Petrov, and H. P. B{\" u}chler, Phys. Rev. A {\bf 98}, 051604(R) (2018).


%in photonic system:
%\bibitem{Valiente}N. Westerberg, K. E. Wilson, C. W. Duncan, D. Faccio, E. M. Wright, P. {\" O}hberg, M. Valiente, arxiv:1801.08539.

%soliton:
%first theory of soliton:
\bibitem{soliton_theory}V. M. P$\acute{e}$rez-Garc$\acute{i}$a, H. Michinel, and H. Herrero, Phys. Rev. A {\bf 57}, 3837 (1998).
%first two expt of bright soliton:
\bibitem{soliton_expt1}K. E. Strecker, G. B. Partridge, A. G. Truscott, and R. G. Hulet, Nature {\bf 417}, 150 (2002).
\bibitem{soliton_expt2}L. Khaykovich, F. Schreck, G. Ferrari, T. Bourdel, J. Cubizolles, L. D. Carr, Y. Castin, and C. Salomon, Science {\bf 296}, 1290 (2002).

%uniform trap:
\bibitem{box_trap1}A. L. Gaunt, T. F. Schmidutz, I. Gotlibovych, R. P. Smith, and Z. Hadzibabic, Phys. Rev. Lett. {\bf 110}, 200406 (2013).
\bibitem{box_trap2}I. Gotlibovych, T. F. Schmidutz, A. L. Gaunt, N. Navon, R. P. Smith, Z. Hadzibabic, Phys. Rev. A {\bf 89}, 061604(R) (2014).
\bibitem{box_trap3}C. Eigen, A. L. Gaunt, A. Suleymanzade, N. Navon, Z. Hadzibabic, R. P. Smith, Phys. Rev. X {\bf 6}, 041058 (2016).
\bibitem{box_trap4}B. Mukherjee, Z. Yan, P. B. Patel, Z. Hadzibabic, T. Yefsah, J. Struck, and M. W. Zwierlein, Phys. Rev. Lett. {\bf 118}, 123401 (2017).
\bibitem{box_trap5}K. Hueck, N. Luick, L. Sobirey, J. Siegl, T. Lompe, and H. Moritz, Phys. Rev. Lett. {\bf 120}, 060402 (2018).

\bibitem{book} C. J. Pethick and H. Smith, {\it Bose-Einstein Condensation in Dilute Gases}, Cambridge University Press, 2002.

%\bibitem{footnote_norm} Here we have ${\cal N}=2\pi^2(\sigma_{xy}\sigma_z)^2/L \sum_{\nu} {\rm exp}(-(\nu \pi\sigma_z/L)^2)$.

%\bibitem{footnote_soliton} We find that the soliton state for small $N$ reproduces the 2D droplet as studied in Ref.\cite{Petrov_2}.

%\bibitem{supple} See supplementary materials for more details.

\bibitem{HuHui}H. Hu and X.-J. Liu, arXiv:2005.08581.
\bibitem{Ota}M. Ota and G. E. Astrakharchik, SciPost Phys. 9, 020 (2020).
\bibitem{YinLan}Q. Gu, L. Yin, arxiv:2005.13154.

\end{thebibliography}
\end{document}